\documentclass[%
 reprint,
 amsmath,
 amssymb,
aip,
jcp,
]{revtex4-1}

\usepackage{graphicx}
\usepackage{dcolumn}
\usepackage{bm}
\usepackage{epsfig,color,xspace,multirow,xr,bbold}
\usepackage[all]{xy}
\usepackage{setspace}
\usepackage{url}
\usepackage{siunitx}

\usepackage{tikz}
\usepackage{xcolor}

\newcommand{\bG}{{\bf G}}

\newcommand{\bphi}{{\boldsymbol \phi}}
\newcommand{\bb}{{\boldsymbol b}}

\usepackage{hyperref} 

\hypersetup{
   colorlinks,
   menucolor=blue,
   linkcolor=black,
   citecolor=blue,
   urlcolor=blue
}

\begin{document}
\title{Coarse-grained conformational surface hopping: Methodology and
transferability}

\author{Joseph F.~Rudzinski}
\email{rudzinski@mpip-mainz.mpg.de}
\affiliation{Max Planck Institute for Polymer Research, 55128 Mainz,
Germany}
\author{Tristan Bereau}
\email{t.bereau@uva.nl}
\affiliation{Van 't Hoff Institute for Molecular Sciences and
	Informatics Institute, University of Amsterdam, Amsterdam 1098 XH, The
Netherlands}
\affiliation{Max Planck Institute for Polymer Research, 55128 Mainz,
Germany}

\date{\today}

\begin{abstract}
	Coarse-grained (CG) conformational surface hopping (SH) adapts the
	concept of multisurface dynamics, initially developed to describe
	electronic transitions in chemical reactions, to accurately
	describe classical molecular dynamics at a reduced level. The SH
	scheme couples distinct conformational basins (states), each
	described by its own force field (surface), resulting in a
	significant improvement of the approximation to the many-body
	potential of mean force [\emph{Phys.~Rev.~Lett.}~\textbf{121},
	256002 (2018)]. The present study first describes CG SH in more
	detail, through both a toy model and a three-bead model of hexane.
	We further extend the methodology to non-bonded interactions and
	report its impact on liquid properties. Finally, we investigate
	the transferability of the surfaces to distinct systems and
	thermodynamic state points, through a simple tuning of the state
	probabilities. In particular, applications to variations in
	temperature and chemical composition show good agreement with
	reference atomistic calculations, introducing a promising
	``weak-transferability regime,'' where CG force fields can be
	shared across thermodynamic and chemical neighborhoods.
\end{abstract}

\maketitle

\section{Introduction}

In the realm of multiscale models for soft matter and biomolecular
systems, particle-based coarse-grained (CG) resolutions have offered
tremendous insight.~\cite{MullerPlathe:2002, Nielsen:2004tw, Voth:2009, Murtola:2009, Peter:2009hr, Peter:2010zc, Ingolfsson:2013, Brini:2013, Noid:2013uq} CG models average over the faster degrees of
freedom by lumping several atoms into super-particles or beads. When adequately 
built and parametrized, these models can strike an
excellent balance between accuracy and computational efficiency. Their
success stems largely from a mapping commensurate with the system's
scale separation and an adequate use of physics-based modeling. The
latter aspect is the main topic of this study. 

Coarse-graining replaces the coveted potential-energy surface (PES) by
a configuration-dependent free-energy function known as the many-body
potential of mean force (MB-PMF).~\cite{Kirkwood:1935ys,Noid:2008a}
Over the last two decades the community has been developing and
improving a number of systematic methods aimed at targeting the
MB-PMF.~\cite{Izvekov:2005e, Lyubartsev:1995, Tschop:1998lj,
Noid:2008a, Shell:2008il, Mullinax:2009b, Brini:2011} While
early efforts established a strong theoretical and practical
foundation for these methods, a number of fundamental challenges have
arisen, which largely prevent a more widespread utilization of
systematic (i.e., bottom-up) CG models.~\cite{Noid:2013uq, Dunn:2016b}
Transferability---the capability of a given model to be accurately
applied to systems and thermodynamic state points distinct from those
used for parametrization---is an intrinsic problem for
coarse-graining, since the MB-PMF is inherently state-point
dependent.~\cite{Riniker:2012qf, Noid:2013uq} As a consequence, there
has been a continued effort to systematically investigate the
temperature, density, and solvent-mixture transferability properties
of CG models.~\cite{Mullinax:2009a, Shen:2011, Rosenberger:2018,
Lebold:2019, Rosenberger:2019, Deichmann:2019, Szukalo:2020} In
limited cases, it has been demonstrated that CG interactions can
reproduce the temperature-dependence of liquid structure through an ad-hoc
linear interpolation,~\cite{Qian:2008,Farah:2011} although a
systematic approach has been lacking. Recent work has begun to fill
this gap either through Bayesian techniques~\cite{Patrone:2016} or
approaches that approximate the entropic contributions to the
effective potentials, allowing explicit predictions of state-point
dependence.~\cite{Lebold:2019b, Lebold:2019c, Jin:2019, Jin:2020} 
These studies
have focused on CG representations without significant
intramolecular flexibility. Beyond thermodynamic-state-point
dependence, few studies have reported detailed characterizations of
the chemical transferability of bottom-up CG models.~\cite{Wang2010,
Brini:2012,Brini:2012b,Ohkuma:2020,Shen:2020}

Even for a single system or thermodynamic state point, persistent
efforts have not led to steady improvements in the quality of the
force fields---the accuracy being limited less by the performance of
the methods, and more by the molecular-mechanics terms used to
approximate the MB-PMF. Because these terms only offer an incomplete
representation of the full MB-PMF, a CG model's accuracy critically
depends on two aspects: ($i$) an optimized mapping that most
effectively simplifies the form of the MB-PMF,~\cite{Rudzinski:2014,
Foley:2015, Chakraborty:2018, Chakraborty:2019, Giulini:2020} and
($ii$) interaction-potential forms that are flexible enough to
describe complex physical phenomena, such as interfaces or
environment-dependent conformational
changes.~\cite{Jochum:2012wa,Dalgicdir:2016} Going beyond the typical
interaction terms---especially non-bonded pairwise interactions---can
have significant impact, as seen by recent investigations that
considered physics beyond pairs, such as three-body
interactions~\cite{Molinero:2009fu, Larini:2010dq, Das:2012c,
Lindsey:2017, Scherer:2018} or local density-dependent
potentials.~\cite{DeLyser:2017, Sanyal:2018, Jin:2018, DeLyser:2019,
Shahidi:2020} However, these approaches are also limited by the
functional forms applied, and it may not be obvious how to generalize
them for optimal improvement in modeling accuracy. Recent applications
using machine learning can provide a more accurate reproduction of the
MB-PMF either through a multi-body decomposition or by a direct
interpolation of the many-body forces.~\cite{John:2017, Zhang:2018,
Wang:2019, Chan:2019, Scherer:2020, Ruza:2020} This improved accuracy
typically comes with added computational cost---a significantly larger
evaluation time needed for the force prediction,~\cite{Zuo:2020} which
can be mitigated with the use of tabulated
potentials.~\cite{Scherer:2020} 

We recently introduced a complementary approach to improve the
description of cross correlations between interaction terms in a force
field.~\cite{Bereau:2018} This approach was inspired by the
modeling of chemical reactions, where distinct electronic
configurations are decomposed onto separate surfaces in order to
overcome limitations of the force field by coupling distinct
PESs---notable examples include (multisurface) empirical valence bond
and surface-hopping schemes.~\cite{Warshel:1980,
Schmitt:1998, Tully:1990} Instead of describing
transitions between electronic states, our method considers switching
between conformational basins: Distinct force fields describe
interactions for a subset of conformational space. 
There has been a number of previous efforts to couple internal states
in various ways within the context of classical molecular simulations.~\cite{Murtola:2009ys, Chappa:2012, Knott:2014, Pagonabarraga:2001, Louis:2002, Allen:2008, Ilie:2016}
These studies have avoided explicit hopping schemes through
approaches that either ($i$) linearly interpolate between two
force fields (e.g., multi-state G\={o} models) or ($ii$) describe the
force-field change as an analytic function of a continuous order
parameter (e.g., local density-dependent potentials).
Voth and coworkers formalized the employment of internal states within
simulation models through a bottom-up ``ultra coarse-graining'' framework,~\cite{Dama:2013} 
originally used to develop models that stochastically and discretely transition
between internal states.~\cite{Davtyan:2014,Katkar:2018}
These studies considered the regime in which there exists a clear
time scale separation for internal state transitions.
The framework was later extended to the regime of ``rapid local equilibrium''~\cite{Dama:2017}---
transitions occur very quickly relative to the translational motion of the CG sites---and
deployed to accurately describe interfacial properties~\cite{Jin:2018b} and hydrogen-bonding~\cite{Jin:2018} in 
molecular liquids while using only single-site representations for the CG molecules.
In contrast, the surface-hopping method considers an intermediate regime, where transitions between 
local conformational basins occur on time scales that are on par with other molecular motions.
As a result, we focus on identifying conformational basins according to {\it intramolecular} CG
degrees of freedom.
Sharp \emph{et al.} recently extended this idea, based on an empirical valence bond perspective,
to describe transitions between conformational basins
defined along a set of collective variables.\cite{Sharp:2019}

Coupling interaction terms of the Hamiltonian offers the possibility
to rescue cross correlations beyond the typical global separation of
variables. By focusing on the coupling of \emph{intramolecular}
interactions, we reported significant improvements in the accuracy of
the approximation to the MB-PMF for a three-bead model of hexane, as compared to the
baseline force-matching-based multiscale-coarse-graining (MS-CG)
method.~\cite{Bereau:2018} Furthermore, the surface-hopping model for a tetra-alanine
peptide in water not only resulted in significant improvements of the 
two-dimensional projection of the MB-PMF but also reproduced (within
error bars) the ratio of mean-first passage times between helical and
extended states. The latter is significant: it shows that a faithful
representation of the MB-PMF can offer an accurate reproduction of the
barrier-crossing dynamics up to a speedup factor. While equilibrium
properties depend exponentially on the free-energy minima, an accurate
reproduction of the barrier-crossing dynamics critically depend on the
free-energy \emph{barriers}.~\cite{Rudzinski:2016, Rudzinski:2016b,
Rudzinski:2018b}

The present report extends our previous work in several ways. We first
provide a more detailed account of the methodology starting from a toy
example---a single particle in a double-well potential. Next, we
extend the methodology to non-bonded pairwise interactions and report
results on liquid properties. Finally, we investigate
``weak-transferability'' properties, corresponding to the transfer of
surfaces while solely tuning the state probabilities (i.e., their
prefactors). We observe a monotonic---almost linear---variation in
state probabilities as a function of temperature and chemical
composition. The results suggest that decomposing CG force fields into
surfaces may facilitate transfer across state-point neighborhoods.

\section{Methodology}

For completeness we first recall the protocol applied to
intramolecular interactions.\cite{Bereau:2018} This is followed by
the extension to intermolecular interactions.

\subsection{Intramolecular interactions}

\begin{figure}[htbp]
	\begin{center}
		\includegraphics[width=0.90\linewidth]{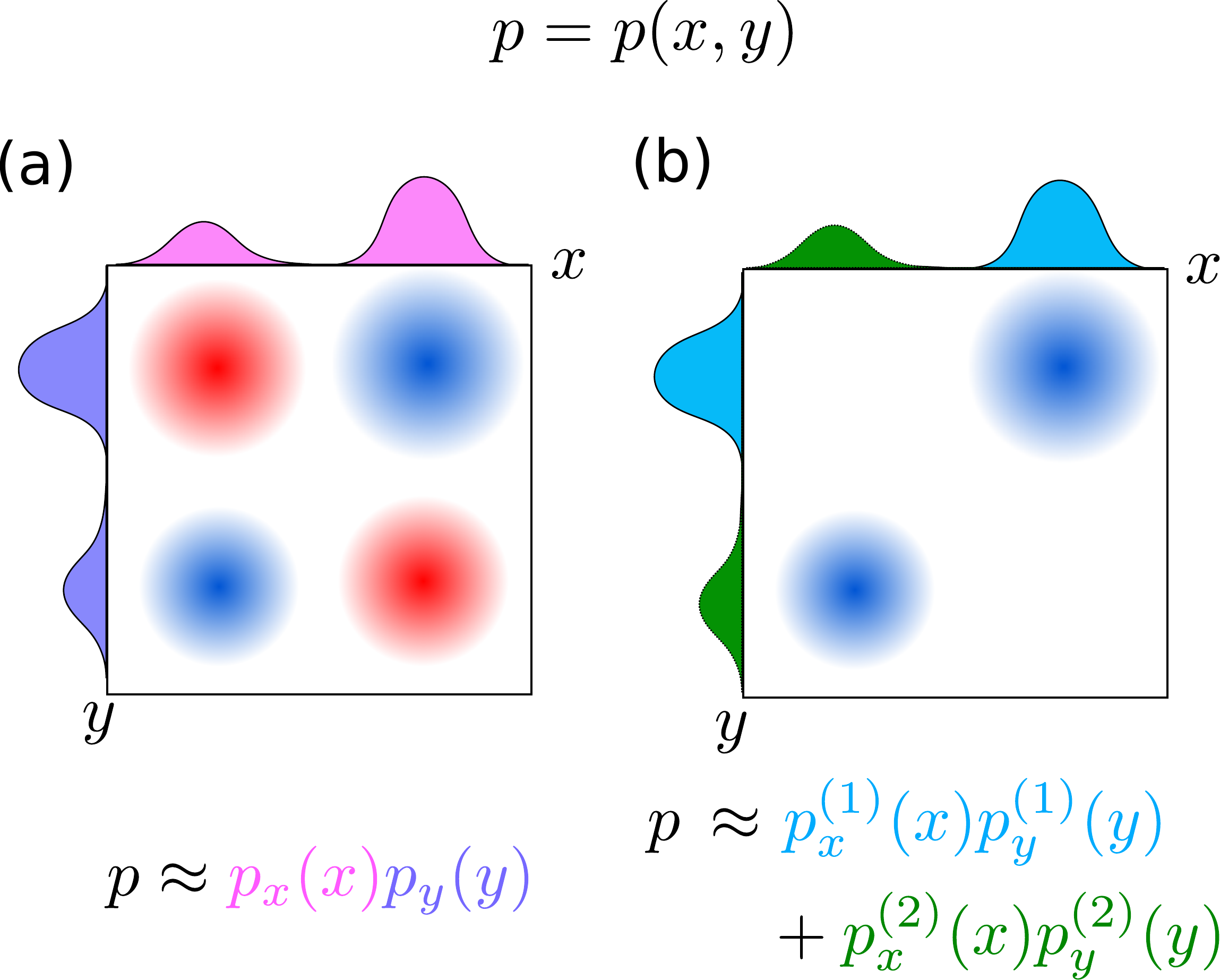}
		\caption{(a) Standard force fields apply a global separation of
			variables on the probability distribution $p=p(x,y)$, leading to
			$p \approx p_x(x) p_y(y)$. (b) Surface hopping, while retaining
			the separation of variables, ascribes one \emph{local} force field
			per conformational basin.  Adapted from
			Ref.~\onlinecite{Bereau:2018}.}
		\label{fig:sketch}
	\end{center}
\end{figure}

We recall the example presented in our previous
publication:~\cite{Bereau:2018} We consider a two-dimensional
potential $U=U(x,y)$ with corresponding canonical equilibrium
distribution $p = p(x,y) \propto \exp\left( - \beta U(x,y)\right)$,
where $\beta= (k_{\rm B}T)^{-1}$ is the inverse temperature.  Standard
molecular-mechanics force fields apply a global separation of
variables on the potential, such that $U(x,y) \approx U_x(x) +
U_y(y)$, also impacting the equilibrium distribution: $p(x,y) \approx
p_x(x) p_y(y)$. As a result, we cannot ensure an accurate reproduction
of cross correlations between $x$ and $y$. For instance, the
intramolecular interactions of a 3-particle, linear molecule made of
two bonds, $b_1$ and $b_2$, and one bending angle $\theta$ will
typically be modeled by a potential of the form $U(b_1, b_2, \theta) =
U_{b_1}(b_1) + U_{b_2}(b_2) + U_\theta (\theta)$. While significantly
advantageous from a computational standpoint, the separation of
variables can drastically hamper the accuracy of the (free-)energy
landscape.  Fig.~\ref{fig:sketch}a illustrates the potential issues of
such an approach. In particular, if there exists two local minima
along each degree of freedom, a model which employs the global
separation of variables will likely sample all four combinations of
these minima, regardless of the true underlying distribution.

The conformational surface-hopping (SH) scheme retains the same form of
the Hamiltonian, as well as the separation of variables, but ascribes
a \emph{local} force field for a subset of conformational space---a
conformational basin, say. In the case of two surfaces, the SH equilibrium
distribution takes the form $p(x,y) \approx p_x^{(1)}(x)
p_y^{(1)}(y) + p_x^{(2)}(x) p_y^{(2)}(y)$, allowing the description of
a wider range of cross correlations between the degrees of freedom
(see Fig.~\ref{fig:sketch}b). This prescription trivially generalizes
to $n$ surfaces. An SH force-field parametrization thereby consists of
the following steps:
\begin{enumerate}
	\item A clustering of (intramolecular) conformational space is performed (here with respect to
	      variables $x$ and $y$) to identify homogeneous
	      regions, ideally leading to unimodal one-dimensional distribution functions
	      along each (intramolecular) degree of freedom. 
	      Each cluster is assigned a center, $\mu^{(i)} = \left( \mu_x^{(i)}, \mu_y^{(i)} \right)$,
	      corresponding to the local maximum of probability density, and a spatial extent, $\sigma^{(i)} =
	      \left(\sigma_x^{(i)}, \sigma_y^{(i)} \right)$, related to the standard deviation of configurations 
	      belonging to the cluster.
	\item A linear transformation is applied to the conformational space in order to enhance the isotropy
	      of the clusters: $\overline\sigma^{(i)} = \left(\overline\sigma_x^{(i)},
	      \overline\sigma_y^{(i)} \right)$.
	\item $n-1$ surfaces are defined according to the clustering, while an additional surface is introduced
	      which covers the remaining configurations.
	      This surface will be referred to as the ``fallback'' surface.
	\item A structure-based parametrization of $n$ force fields is performed (e.g., via force matching), 
	      one for each surface.
\end{enumerate}

Each SH force field, ${\bf f}_i({\bf R}) = -\nabla U_i({\bf R})$, is related to a typical molecular mechanics potential, $U_i({\bf R})$, which employs a global separation of variables.
In the SH method, the net force for any configuration of the system can be written as a linear combination of the individual force fields:
\begin{equation}
	{\bf f}({\bf R}) = \sum_{i=1}^n w_i {\bf f}_i ({\bf R}),
\end{equation}
where the coefficients or weights are restricted to $0 < w_i < 1$.
Force field $i$ will contribute to the net force
according to the proximity of the system's instantaneous configuration
to cluster $i$. Practically $w_i$ is computed as a Euclidean distance
of the system's CG interaction variables $(x,y)$ to the cluster center
\begin{equation}
	d_i = \sqrt{\frac{(x-\mu_x^{(i)})^2}{\overline\sigma_x^{(i)}} 
		+ \frac{(y-\mu_y^{(i)})^2}{\overline\sigma_y^{(i)}}}.
\end{equation}
$d_i$ is then compared to the spatial extent of the cluster
$|\overline\sigma^{(i)}|$. When $d_i < |\overline\sigma^{(i)}|$, the
system is completely within cluster $i$ and its force field receives the full
weight, $w_i = 1$, while all other force fields are neglected. In the case that the system's
configuration is \emph{not} inside one of the clusters, the SH approach will
connect surfaces together to ensure a smooth hopping. To this end, the force field weight is
exponentially suppressed with respect to the distance from
the \emph{boundary} of the cluster, $d_i - |\overline{\bf
	\sigma}^{(i)}|$:
\begin{equation}
	\label{eq:wi}
	w_i =
	\begin{cases}
		1 , \quad & d_i < |\overline{\bf \sigma}^{(i)}| \\
		\exp \left(- \frac{d_i - |\overline{\bf
		\sigma}^{(i)}|}\alpha \right) , \quad
		          & \textup{otherwise}.                 
	\end{cases}
\end{equation}
The sharpness of this suppression is determined via the scaling parameter $\alpha$,
which can assist in avoiding numerical
instabilities in the simulations. On the other hand, we stress the
importance of keeping $\alpha$ small, as it blurs the force-field
boundaries. 

Mixing different force fields can lead to unphysical behavior, for
instance if the aggregate contributions yield large net forces.  This
is especially relevant at the boundaries between conformational
basins, where a localized force field will have large restoring forces
at the boundaries (see, for instance, panels a and b of
Fig.~\ref{fig:toy}). We hinder this behavior by restricting mixing to
occur between only two force fields: one corresponding to the closest
cluster and one corresponding to the fallback surface. More
specifically, we first compute the initial $w_i$ for each of the first
$n-1$ surfaces according to Eq.~\ref{eq:wi}. The largest weight, $w_l
= \max_{i<n} w_i$, is kept, while the remaining weights are set to
zero. Then, the final weight is assigned to the fallback surface: $w_n
= 1-w_l$. This approach assumes that the fallback surface is well
connected to all of the surfaces and, consequently, is well-defined
broadly across the conformational space of the system. Akin to
force-based adaptive resolution simulations, the present protocol can
lead to non-conservative forces, requiring the use of a local
thermostat (e.g., via Langevin dynamics; see also Fig.~S1 in the
Supporting Information (SI)).~\cite{Kreis:2014} 

As described thus far, the algorithm leads to surface hopping but does not ensure the
correct probability of sampling each conformational basin. To this
end, we enforce that the time average of the probabilities to be
within each state roughly matches a set of target reference
probabilities, available upon partitioning conformational space. This
approach, both simple to implement and effective, is described in more
detail in our previous work,~\cite{Bereau:2018} as well as below.

\subsection{Intermolecular interactions}

Having described an SH model that switches between force fields
according to the order parameters governing intramolecular
interactions, we now turn to the treatment of intermolecular
interactions. In this work, the intermolecular interactions rely on
the SH state definition, determined by the intramolecular order
parameters, which effectively couples the two types of interactions.
However, the local (non-bonded) environment of each molecule does not
play a role in defining the SH state. For instance, consider two
particles of type A and B belonging to distinct molecules. For each
particle we compute their most contributing surface---a function of,
for example, the bond distances and bending angles of each molecule.
Let these surfaces be denoted $j$ and $k$ for particle types A and B,
respectively. Then, the resulting pairwise non-bonded interaction
between these particles will not only depend on the pair of particle
types A--B, but \emph{also} on the combination $j$--$k$. The
parametrization of the intermolecular interactions consists of
appropriate filtering of the reference trajectory: we gather statistics
between particles A and B that also have internal state $j$ and $k$,
respectively. Computationally, the non-bonded interaction switches
nearly instantaneously according to the pair of internal states, as
defined by the bonded interactions. The relatively small difference
between non-bonded potentials helps avoid numerical instabilities. 

\section{Computational methods}

The protocol applied here largely follows our previous
study.~\cite{Bereau:2018}

\subsection{All-atom simulations}

In this work, we consider four small molecules: hexane, octane,
hexanediamine, and hexanediol. For each molecule, we performed
simulations of ($i$) a single molecule in vacuum and ($ii$) 267
molecules in the liquid phase at various temperatures. These
simulations employed the OPLS-AA force field~\cite{Jorgensen:1996} to
model interactions and were performed with the Gromacs 4.5.3
simulation suite~\cite{Hess:2008} according to standard procedures,
described in more detail in the SI.

\subsection{Coarse-grained representation and interactions}

For hexane, we considered a 3-site representation, which represents
subsequent pairs of carbon atoms with a CG site. The CG potential
included two identical bonded interactions between subsequent pairs of
sites along the chain and an angle bending interaction between the
three CG sites. This representation and set of interactions has been
applied in several previous
studies.~\cite{Ruhle:2009wx,Ruhle:2011,Rudzinski:2014,Bereau:2018}
Octane, hexanediamine, and hexanediol can be considered ``extensions''
to the hexane molecule, through the addition of a functional group on
each end of the molecule. To assess to what extent the SH state
definitions can be transferred between molecules, we employed the
hexane mapping and interaction set to these other three molecules.
That is, each pair of carbon atoms were represented by a CG site,
while the terminal functional groups were not explicitly represented
(see Fig.~\ref{fig:twodimtransf}(d)). For each of these 3-site models,
the terminal CG sites were represented by identical types, denoted as
CT. The center CG site was represented with a distinct type, denoted
CM. We considered both the case in which pairwise non-bonded interactions were
transferred between molecules and also the case in which 
distinct interactions were employed
between each unique pair of bead types.

\subsection{Partitioning of conformational space}

To obtain the SH state definitions, we performed a density-based
clustering analysis~\cite{Sittel:2016} to the atomistic
trajectories of single molecules in vacuum, after mapping each
configuration to the CG representation. This clustering analysis was
performed along the order parameters governing the intramolecular CG
interactions, i.e., the two bond distances and bending angle.
Before clustering, these intramolecular order parameters
were transformed to mean-centered and normalized values for
regularity. The clustering used a search radius $R=0.1$. The initial
clusters were grouped into coarser states manually via visualization
of the cluster distributions along each order parameter, although an
automated dynamics-based algorithm~\cite{Jain:2012} yielded similar results.

For the three-bead representations of both hexane and octane, the
clustering resulted in a set of 7 clusters, representing different
combinations of bond and bending-angle values.
This is consistent with a previous analysis of the intramolecular
conformations of molecules in liquid hexane,~\cite{Rudzinski:2014}
which showed that the 6 possible dihedral states in the AA representation
(3 dihedrals times 2 possible states each, trans or gauche) are mapped
to 7 CG intramolecular states.
This result already indicates
some consistency between the intramolecular states sampled by these
distinct molecules. In the following, we will consider a 3-state SH
model, where the two most populated clusters (representing $\approx 60\%$ of 
the intramolecular conformations) determined the states
denoted 3S-1 and 3S-2, while the rest of the configurations were
lumped into the fallback surface (3S-3). 
Our previous work demonstrated that this 3-surface representation was sufficient to nearly
quantitatively reproduce the bond-angle cross correlations of hexane in vacuum.~\cite{Bereau:2018}
In the Results section, we
first assess the properties of the SH model for hexane, both in vacuum
and in the liquid phase, using the state definitions determined from
the AA simulations of hexane. Subsequently, the 
transferability of the SH state definitions across chemistry is assessed.
For this investigation, a single set of state definitions, determined from
the AA simulations of octane, were applied for each molecule.

\subsection{Generation of the coarse-grained potentials}

In this work, all CG force fields are derived using the framework of
the force-matching-based multiscale coarse-graining (MS-CG) method.
The MS-CG method approximates the MB-PMF via a mathematical projection of the
many-body mean force, i.e., the negative gradient of the MB-PMF, into
the space of force fields spanned by the chosen basis set
representation for the CG force
field.~\cite{Izvekov:2005e,Noid:2008a} This corresponds to
matching the average force on each CG particle sampled in the
simulation of the underlying, higher-resolution model.  Practically,
the projection is expressed as a linear least squares problem in the
basis function coefficients, i.e., the CG force-field parameters,
$\bphi$, and can be written in the normal equation representation as
\begin{equation}
	\label{eq-MSCG}
	\bb^{\rm AA} = \bG^{\rm AA}\bphi^{\operatorname{MS-CG}}      .
\end{equation}
In Eq.~\ref{eq-MSCG}, $\bb^{\rm AA}$ is a vector of ensemble averages
that can be expressed as a set of either force~\cite{Noid:2007a,
	Noid:2008a} or structural~\cite{Mullinax:2009b, Mullinax:2010,
Ellis:2011a} correlation functions. The latter is possible through a
generalized Yvon-Born-Green (g-YBG) framework, which connects the
MS-CG method to traditional liquid state
theory.~\cite{Mullinax:2009b,Mullinax:2010} For a non-bonded, pairwise
interaction represented by a set of spline basis functions, $\bb^{\rm
AA}$ is directly related to the corresponding radial distribution
function generated by the reference model but mapped
to the CG representation.~\cite{Ellis:2011a} $\bG^{\rm AA}$ is a matrix
that quantifies the cross correlations between pairs of CG degrees of
freedom generated by the reference model.  If the model
derived from the MS-CG method fails to reproduce the target vector of
these equations, i.e., $\bb^{\rm AA}$, it implies that the
cross-correlation matrix generated by the higher resolution model does
not accurately represent the correlations that would be generated by
the resulting CG model.  This indicates a fundamental limitation of
the model representation and interaction set.  Nevertheless, the
system of equations can be solved self-consistently to determine the
force field $\bphi^*$ that reproduces the target vector, albeit at the
expense of the representation of the cross correlations of the
underlying model~\cite{Rudzinski:2014}:
\begin{equation}
	\label{eq-itergYBG}
	\bb^{AA} = \bG(\bphi^*) \bphi^* .
\end{equation}
This approach has been previously denoted as an iterative g-YBG
(iter-gYBG) method.~\cite{Cho:2009ve,Lu:2013uq,Rudzinski:2014} In the
following, we consider both $\bphi^{\operatorname{MS-CG}}$ and
$\bphi^*$ (denoted as the iter-gYBG model), simulated according to
standard techniques, for comparison with the SH simulation method
described above.

Using the partitioning of configuration space described above, we also
determine sub-ensemble-specific CG potentials by solving
Eq.~\ref{eq-MSCG}, but employing trajectories containing only
configurations from a specific sub-ensemble to calculate each of the
correlation functions. Although structure-based methods are often
applied to conformational ensembles at equilibrium, several studies
have demonstrated the benefit of performing parametrizations over
sub-ensembles or biased
ensembles.~\cite{Brini:2011,Mukherjee:2012wx,Rudzinski:2014b,Shen:2020} The
formal theory for such calculations in the context of the MS-CG method
has been detailed by Voth and coworkers.~\cite{Dama:2013,Davtyan:2014}
All force-field calculations in this work were performed using the
BOCS package.~\cite{Dunn:2018} Further numerical details for these
calculations are provided in the SI. While the main
text compares force fields via graphical or qualitative means, the SI
contains a quantitative comparison of the accuracy of the force fields
by means of the Jensen-Shannon divergence.

\subsection{Coarse-grained simulations}

We performed CG simulations of the SH models using a modified version
of {\sc ESPResSo++}.~\cite{Halverson:2013} Simulations in the
canonical ($NVT$) ensemble were performed using a Langevin thermostat
at various temperatures (more details below and in the SI), where a
friction constant $\Gamma = 10\,\tau^{-1}$, was applied. Here, $\tau$
corresponds to the intrinsic unit of time of the CG model. We
integrated the equations of motion with a time-step $\delta t =
0.001\,\tau$.
All cluster sizes $\{\sigma_i \}$ were scaled by a factor of 0.4 to
significantly localize each surface. The smoothness scaling parameter
was set to a small value, $\alpha = 0.05$, to ensure numerical
stability of the dynamics while minimally distorting the individual
force fields.
An {\sc ESPResSo++} implementation of the CG surface-hopping scheme,
including support for non-bonded interaction, is available
online.~\cite{gh_esp}

We performed CG simulations of the MS-CG and iter-gYBG models using
version 4.5.3 of the GROMACS package,\cite{Hess:2008} according to
standard procedures (see SI).

\section{Example: Toy model}

\begin{figure*}[htbp]
	\begin{center}
		\includegraphics[width=0.95\linewidth]{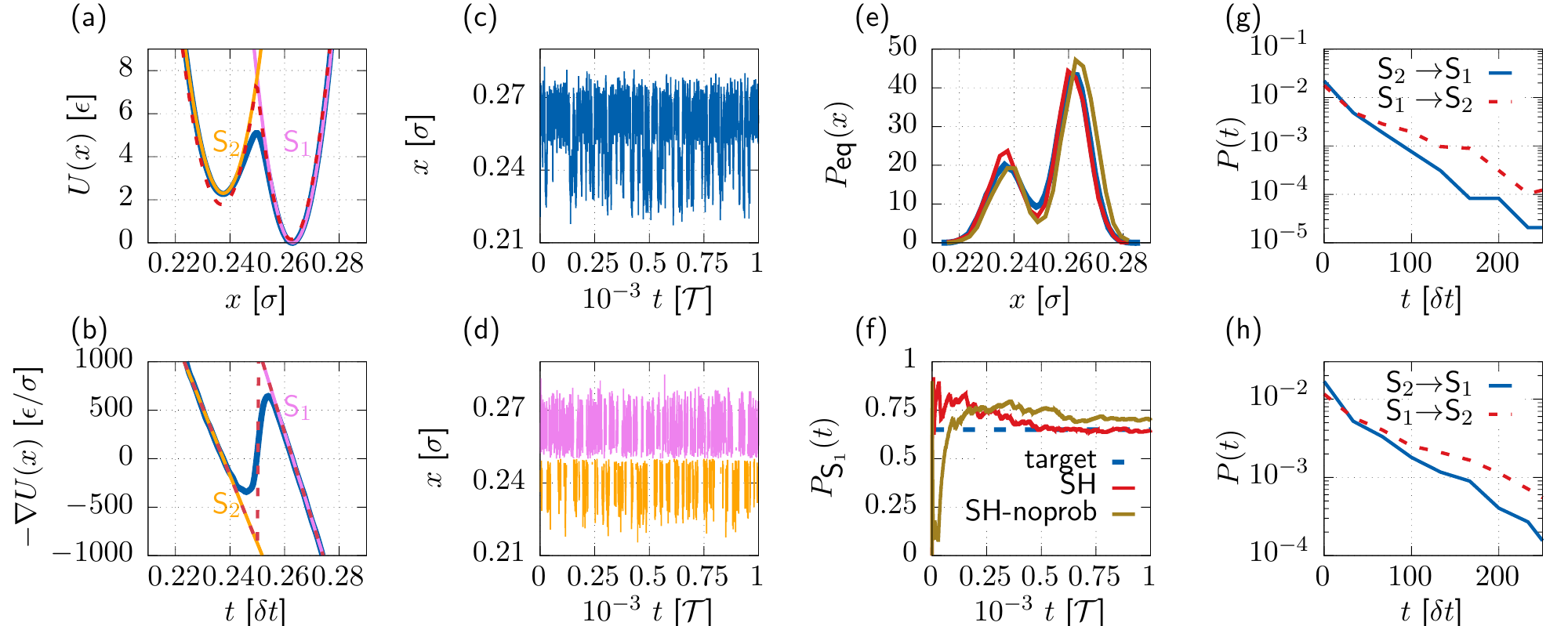}
		\caption{(a) Potential energy $U(x)$ and (b) force $-\nabla
			U(x)$ of the toy model. The global surface is shown in
			thick blue, while the two surfaces S$_1$ and S$_2$ are
			alternatively described by harmonic potentials. Time
			trajectories under Brownian dynamics over (c) the global
			surface and (d) surface hopping between S$_1$ and S$_2$.
			(e) Boltzmann distributions of the reference model (thick
			blue) as well as the surface-hopping model (SH) and the
			alternative without matching probability (``SH-noprob'').
			See also (f) for the color-coding. (f) Time evolution of
			the probability of populating S$_1$ for SH and SH-noprob
			as compared to the reference probability obtained from the
			global surface. (g-h) Probability distribution of escape
			times of the toy model between basins S$_1$ and S$_2$ for
			(g) the global surface and (h) surface hopping. While the
			characteristic time scales are different between the
			models, the \emph{ratio} of characteristic time scales is
		conserved.}
		\label{fig:toy}
	\end{center}
\end{figure*}

We first illustrate the method using a toy model: a single particle
dynamically evolving in a one-dimensional double-well
potential.\footnote{The attentive reader will realize that this
	potential reproduces the bond distribution of the CG hexane described
later in this article.} The potential $U(x)$, which we will refer to
as the global surface, is shown in Fig.~\ref{fig:toy}a, while panel
(b) displays the corresponding force $-\nabla U(x)$.  Throughout this section we
express the results in the natural units of the model: energy
($\epsilon$), length ($\sigma$), mass ($\mathcal{M}$), and time
($\mathcal{T} = \sigma \sqrt{\mathcal{M}/\epsilon}$). The system is
simulated using Brownian dynamics according to the stochastic
differential equation
\begin{equation}
	\label{eq:brownian}
	\frac{{\rm d}x}{{\rm d}t} = - \frac{D}{k_{\rm B}T}\nabla U(x) + \sqrt{2 D} R(t),
\end{equation}
where $R(t)$ is a white-noise process, $T = \epsilon / k_{\rm B}$ is
the temperature of the system, and $D = 10^2~\sigma^2/\mathcal{T}$ is
the diffusion constant.  We use an integration time-step $\delta t =
10^{-7}~\mathcal{T}$.
Integration of Eq.~\ref{eq:brownian} leads to a time trajectory of the
coordinate, $x(t)$ in Fig.~\ref{fig:toy}c, and an equilibrium
distribution, $P_{\rm eq}(x)$ in Fig.~\ref{fig:toy}e, featuring the
expected two peaks.

We now turn to a surface-hopping model of the system. We split the
global surface into two components: a surface corresponding to the global minimum S$_1$ (violet curves in
panels (a) and (b) of Fig.~\ref{fig:toy}) and a higher-energy surface S$_2$ (orange curves). 
Two distinct potentials are fitted to best reproduce the local basins of $U(x)$ within a
harmonic approximation.  The resulting energy functions, and also the corresponding force curves, show high
fidelity to all parts of the global surface except around the barrier
($x \approx 0.25~\sigma$). We connect the two surfaces S$_1$ and S$_2$
by means of an instantaneous switching at $x=0.25$, leading to a
discontinuity in the force (red dashed line in Fig.~\ref{fig:toy}b).
This generates a cusp in the potential energy, leading to inaccuracies
in the shape of the potential energy around the barrier. A
straightforward integration of the equations of motion of this
surface-hopping model (denoted ``SH-noprob'') qualitatively samples
the two surfaces by regularly switching between them (Fig.~\ref{fig:toy}d), but leads
to noticeable discrepancies in the equilibrium distribution. 
Fig.~\ref{fig:toy}e demonstrates that the SH-noprob simulation slightly overpopulates S$_1$. 
This overrepresented sampling of S$_1$ is clearly displayed in the time evolution of the probability of that
surface (Fig.~\ref{fig:toy}f), which converges to around $P_{{\rm
	S}_1}(t\rightarrow \infty) = 0.70$ instead of 0.65.

A correction to the inaccurate representation of the barrier can be 
obtained by enforcing the probability of sampling S$_1$. To this end,
we restrict the hopping between surfaces by adjusting the force interpolation 
scheme based on the instantaneous time average of the probability of sampling
each surface in the simulation.
More specifically, once the system completely enters a cluster, the weight given to the 
corresponding force field is fixed to be 1 until the probability of sampling the cluster
exceeds a given target probability.~\cite{Bereau:2018}
The surface-hopping simulations with this restriction, denoted simply ``SH,'' 
converges by construction to the target probability
$P_{{\rm S}_1}(t\rightarrow \infty) = 0.65$, leading to an improved
description of equilibrium distribution (Fig.~\ref{fig:toy}e). Thus,
enforcing the target probabilities mitigates potential issues due to
an inaccurate modeling of the boundaries between surfaces. We
emphasize the need for a small interpolation regime between surfaces:
too large of a region would lead to the inclusion of unreasonably large
forces from the less favored surface, resulting in artifacts 
at the interface. We also note that an alternative approach could
consist of interpolating between potential energies, although this
would require to shift each surface by an appropriate amount.

To further probe the dynamics, Fig.~\ref{fig:toy} (g-h) presents the
probability distribution of escape times between basins S$_1$ and
S$_2$. Assuming single-exponential kinetics, we focus on the
characteristic time scales of the forward and backward processes,
$k_{\textup{S}_1\rightarrow\textup{S}_2}$ and
$k_{\textup{S}_2\rightarrow\textup{S}_1}$, respectively. While the
integration of the global surface and the surface-hopping surfaces
(panels a and b, respectively) lead to different characteristic time
scales, their \emph{ratio} are similar:
$k_{\textup{S}_1\rightarrow\textup{S}_2} /
k_{\textup{S}_2\rightarrow\textup{S}_1} \approx  1.47$ and $1.55$,
respectively. This is on par with our previous conclusions about the
method's capability to conserve the \emph{barrier-crossing dynamics},
as illustrated on a tetraalanine peptide.~\cite{Bereau:2018}

\section{Results}

\subsection{Hexane}

In the following we consider the coarse-graining of hexane to a
three-bead representation. We first simulate a single molecule in
vacuum, effectively focusing on the \emph{intra}molecular
interactions.  Later we turn to \emph{inter}molecular interactions by
probing the liquid state.

\subsubsection{Hexane in vacuum}
\label{sec:hex_vac}

The modeling of hexane in vacuum using a 3-site CG representation, 
though presumably straightforward
at first sight, displays remarkably rich cross correlations between
the bond and bending angle degrees of freedom.
This offers a stringent test for
molecular mechanics force fields. The system was first studied by
R\"uhle \emph{et al.}~\cite{Ruhle:2009wx} using the
force-matching based multiscale coarse-graining (MS-CG) method and
later by Rudzinski and Noid, focusing on the cross correlations and
presenting results based on the iterative generalized Yvon-Born-Green
(iter-gYBG) scheme.~\cite{Rudzinski:2014} Some of the analysis
presented here was described in previous
work,~\cite{Bereau:2018} although the present work provides additional details and
uses the previous analysis as a basis to dive further into various features of the method.

\begin{figure}[htbp]
	\begin{center}
		\includegraphics[width=0.90\linewidth]
		{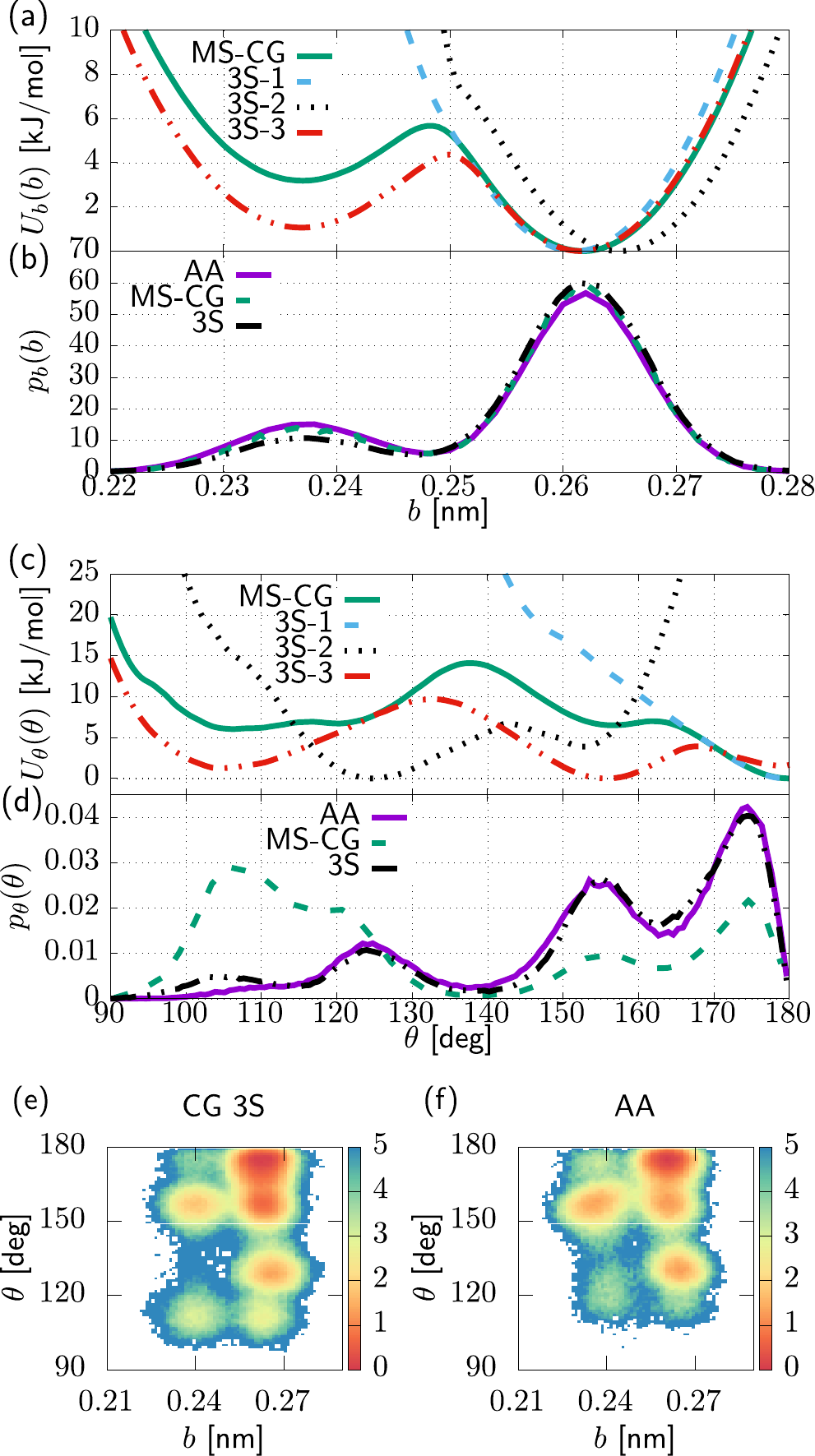} \caption{Bond and
			bending-angle properties of the CG hexane molecule from
			the reference all-atom distribution projected onto the CG
			variables (AA), force matching (MS-CG) and the three-state
			SH model (3S). Bond (a) potential energy and (b)
			probability distribution; and angle (c) potential energy
			and (d) probability distribution. (e-f) Free-energy
			surfaces of the hexane molecule as a function of the bond,
			$b$, and bending angle, $\theta$, from (e) the CG
			three-state surface hopping and (f) reference AA. Free
			energies expressed in $k_{\rm B} T$. Adapted from Bereau
			and Rudzinski.~\cite{Bereau:2018}}
		\label{fig:hexvac}
	\end{center}
\end{figure}

To build the SH model of hexane, we first partitioned the
conformational space defined by the two order parameters governing CG
interactions: bond, $b$, and bending angle, $\theta$. The torsional
degrees of freedom at the atomistic level give rise to a bimodal
distributions of CG bond distances and an approximately trimodal distribution of
angles (violet curves in Fig.~\ref{fig:hexvac}). The angle
distribution also displays a tail at short distances, which
corresponds to a partially hidden fourth mode, described further
below. By separating each order parameter into distinct states based
on these distributions, the intramolecular state of the molecule can
be described as a discrete set of two bond states and an angle state.
The AA model then samples approximately six unique intramolecular
states, with varying equilibrium probabilities.~\cite{Rudzinski:2014}
The surface-hopping model simplifies this description with a 3-state
representation for the intramolecular configuration of the hexane
molecule. This leads to the definition of three surfaces denoted 3S-1,
3S-2, and 3S-3, which we will characterize below. Notably, an analysis
of the reference AA simulation 
provides the probability of sampling
each surface: 0.45, 0.14, and 0.41, respectively.

Fig.~\ref{fig:hexvac} shows both the potential energy and resulting
distribution functions for the bond, $b$, and bending angle, $\theta$,
from the reference all-atom model projected onto the CG variables
(AA), force matching (MS-CG), and the three-state SH model (3S).  Panels
a and b of Fig.~\ref{fig:hexvac} show that the MS-CG model is capable of
reproducing the bond distribution, characterized by a short bond ($b
\approx 0.24$~nm) and a long bond ($b \approx 0.26$~nm). The 3S model
generates essentially 
the same distribution, interestingly using a rich
combination of bond potentials (Fig.~\ref{fig:hexvac}a).  While 3S-1
is dedicated to describing the long bond, 3S-2 is shifted to values
that are even larger---the small probability of sampling this surface
leads to a virtually negligible impact on the bond probability
distribution. Lastly, 3S-3 describes both the long bond---with a basin
aligned with 3S-1---and a short bond that is energetically offset.
This surface alone is responsible for the smaller peak in
Fig.~\ref{fig:hexvac}b.

Turning to the bending-angle potential and probability distribution
(panels c and d of Fig.~\ref{fig:hexvac}), the MS-CG model displays severe
discrepancies: it significantly under-samples the two larger angles
($\theta \approx 170^\circ$ and $\theta \approx 155^\circ$) and
over-stabilizes the two lower angles ($\theta \approx 105^\circ$ and
$\theta \approx 125^\circ$). This discrepancy has been demonstrated to
be due to complex AA cross correlations between the bond and angle
degrees of freedom, which are used as a proxy for CG correlations
within the MS-CG procedure.~\cite{Rudzinski:2014} Unlike the MS-CG model, the
iter-gYBG model presented by Rudzinski and Noid is capable of
reproducing the one-dimensional distribution function
$p_\theta(\theta)$, but does not accurately reproduce the cross
correlations $p(b, \theta)$.~\cite{Bereau:2018} The 3S model
also matches the AA bending-angle distribution nearly quantitatively, but describes
the sub-populations of the distribution with greater detail through
the multi-surface representation. The 3S-1 surface focuses solely on
the largest-angle state, while 3S-2 focuses on the two intermediate
angles.  We note that despite the predominance of the lower
intermediate angle ($\theta \approx 125^\circ$) within 3S-2, the
higher intermediate angle displays a higher population due to 3S-3.
The 3S-3 fallback surface does not target a particular conformational
basin, but instead broadly covers the entire dynamic range of
populated angles with various weights.

The major improvements of the SH model can be seen through the
cross correlations, namely the free-energy surface $-k_{\rm B}T \ln
p(b,\theta)$, displayed in Fig.~\ref{fig:hexvac} (e-f). We previously
showed that the iter-gYBG model yields exceedingly symmetric
features, illustrative of the additivity of the
interactions.~\cite{Bereau:2018} On the other hand, the 3-state SH
model displays a much more accurate free-energy surface.

\subsubsection{Liquid hexane}
\label{liq-hex}

We now turn to assessing the capabilities of the SH models to
describe liquid properties. As a test system we focus on a
homogeneous bulk liquid of hexane, comprised of 267 molecules in a
cubic box of size $L=3.89$~nm\xspace simulated at $T=300$\,K. In
principle, the surface definitions could be extended to depend on
additional order parameters, e.g., as a function of local density.
However, since the benefit of local density-dependent potentials has
already been characterized by others,~\cite{DeLyser:2017,
Sanyal:2018, Jin:2018, DeLyser:2019, Shahidi:2020} here we
focus on the extent to which a more accurate treatment of
intramolecular structure impacts the resulting properties of the
liquid. To this aim, we employ the surface definitions derived from
the vacuum case, described above. While the surface definition {\it
does not} depend on the intermolecular environment, the intermolecular
interactions {\it do} depend on the intramolecular state of the
molecule. That is, we calculate distinct pairwise interactions as a
function not only of the set of bead types but also as a function of
the surface of each molecule. As an illustrative example, we focus on
the interactions between terminal beads, CT, but additional results
for the other interactions can be found in the SI.

\begin{figure*}[htbp]
	\begin{center}
		\includegraphics[width=0.90\linewidth]
		{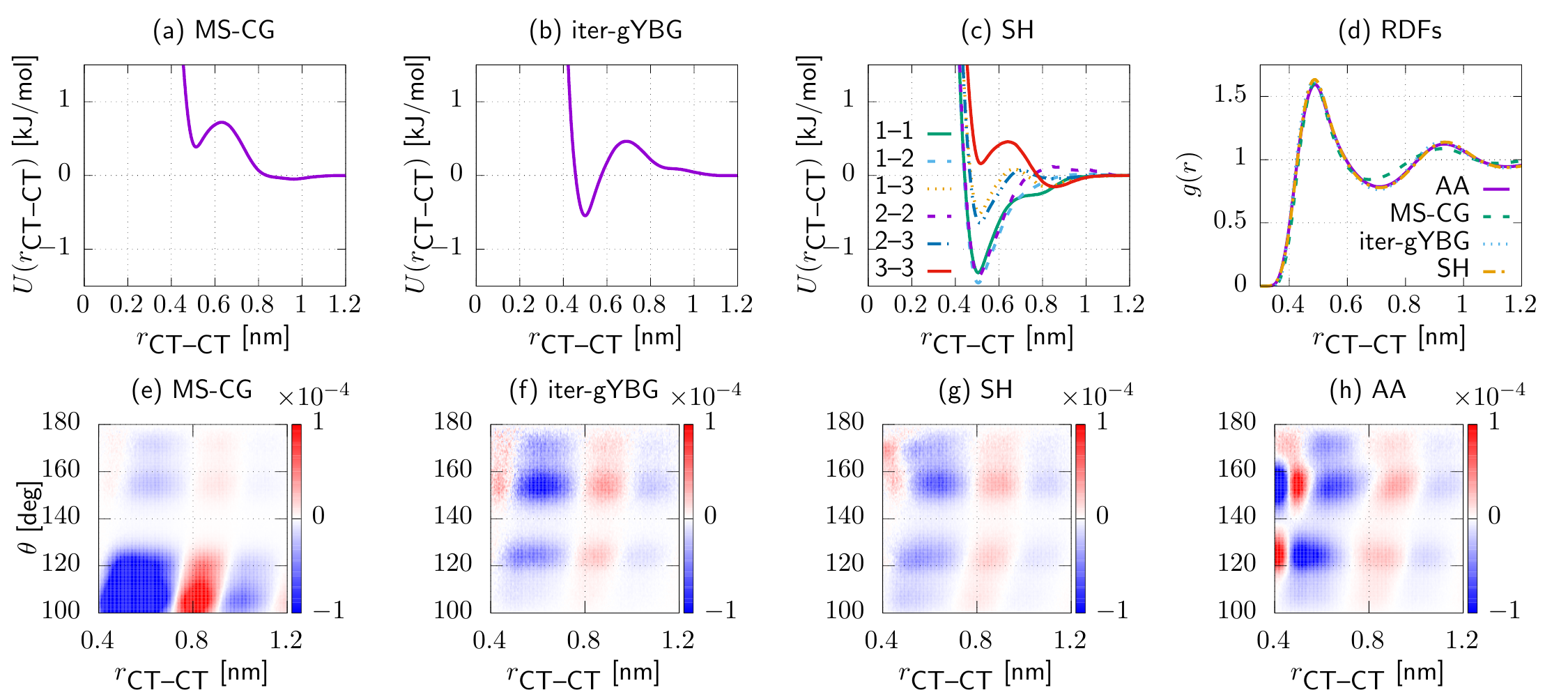} \caption{Hexane liquid
			properties. CT--CT non-bonded potentials from (a) MS-CG, (b)
			iter-gYBG, and (c) SH. The latter shows each surface individually.
			(d) CT--CT radial distribution functions across the three CG
			methods and AA. Cross-correlations between bending angle $\theta$
			and intermolecular distance $r_\textup{CT--CT}$ from (e) MS-CG,
		(f) iter-gYBG, (g) SH, and (h) AA.}
		\label{fig:liqhexane}
	\end{center}
\end{figure*}

Fig.~\ref{fig:liqhexane} (a--c) shows a comparison of the pairwise
interaction potential $U(r_\textup{{CT--CT}})$ for the MS-CG,
iter-gYBG, and SH models. The MS-CG and iter-gYBG potentials are
purely repulsive and only very weakly attractive, respectively,
consistent with a variety of work which has demonstrated that
structure-based methods tend to underestimate the cohesive energy of
liquids.~\cite{Johnson2007,Wang:2009ol,Guenza:2015,Dunn:2016b}
Interestingly, while the SH potentials do include some more repulsive
interactions on par with the MS-CG and iter-gYBG potentials, there are
also some significantly more attractive interactions. The 3--3 SH
potential between fallback surfaces roughly resembles the MS-CG
potential, although it displays an additional small distant attractive
basin around $r_\textup{CT--CT} \approx 0.8$~nm. On the other hand,
the 1--3 and 2--3 SH potentials show a dip that is akin to the
iter-gYBG model, with a depth of about 0.5~kJ/mol, albeit without a
large barrier around $r \approx 0.7$~nm. This is quite striking, since
such barriers and secondary potential minima have been associated to a
type of over-fitting that occurs in structure-based
models.~\cite{Potter:2019} Further, the 1--1, 1--2, and 2--2 SH
potentials show a deeper minimum (1.3~kJ/mol).
This results in a significant reduction in the average pressure throughout 
the ($NVT$) simulation, as seen in Table~\ref{tab:press}, although the SH
models still overestimate the cohesive energy.
We found that this reduced pressure effect occurs systematically for SH models
constructed for three other chemistries (Table~\ref{tab:press}, discussed further below and in the SI).
Critically, we emphasize that there is a clear clustering of the set of SH potentials
into three families: ($i$) the interaction between two molecules both
in the fallback surface (3--3), ($ii$) interactions of a molecule in
the fallback surface with a molecule in one of the other two surfaces
(1--3 and 2--3), and ($iii$) interactions between two molecules not in
the fallback surface (1--1, 1--2, and 2--2). Natural groupings such as
these provide a clear strategy for addressing the combinatorial
explosion of the SH framework as the number of surfaces and bead types
increase.

\begin{table}[htbp]
	\begin{ruledtabular}
		\begin{tabular}{l|r|rrr}
			& $T$ [K] & \multicolumn{3}{c}{$P$ (kbar)}    \\
			              &     & MS-CG   & iter-gYBG & SH      \\
			\hline
			Hexane        & 300 & 2.277 & 1.217   & 0.659 \\ 
			Octane        & 350 & 2.540 & 2.550   & 1.500\\ 
			Hexanediamine & 435 & 2.992 & 2.249   & 1.629 \\ 
			Hexanediol    & 470 & 4.389 & 3.765   & 1.501 \\ 
		\end{tabular}
	\end{ruledtabular}
	\caption{Average pressure, $P$, from $NVT$ simulations. For
		consistency, SH model results correspond to chemistry-specific
		models for each molecule, presented in detail in the SI.}
	\label{tab:press}
\end{table}

Fig.~\ref{fig:liqhexane}d presents the CT--CT radial distribution
functions (RDFs) generated by the various models, demonstrating that
calculating the pairwise interactions as a function of intramolecular
state within the force-matching framework is robust (i.e., does not
result in errors in the structural properties). In fact, the SH model
actually demonstrates an improvement with respect to the MS-CG model,
which shows small deviations after the first solvation shell. These
deviations are, of course, at least partially associated with the
inaccurate determination of the MS-CG angle potential.
Fig.~\ref{fig:liqhexane} (e-h) further characterizes the structural
properties of the models via the cross correlations between the
bending angle and pairwise distance $r_\textup{CT--CT}$. These cross
correlations correspond to sub-blocks of the correlation matrix
employed in Eqs.~\ref{eq-MSCG} and \ref{eq-itergYBG}, and are
described in detail elsewhere.~\cite{Rudzinski:2012vn} Compared to the
AA cross correlations (panel h), the cross correlations generated by
the MS-CG model (panel e) demonstrate significant discrepancies,
largely due to the inaccurate description of the bending-angle
distribution (Fig.~\ref{fig:hexvac}d). In contrast, the cross
correlations generated by the iter-gYBG and SH models demonstrate
better agreement with the AA model, despite some discrepancies at very
short distances ($r \approx 0.4$~nm). 
The intramolecular cross correlations demonstrate analogous behavior 
as in vacuum, with the SH model exhibiting a significantly improved 
representation of the bond-angle correlations (Fig.~S6).
While traditional molecular mechanics potentials fail to describe the
intramolecular cross correlations (Fig.~\ref{fig:hexvac}),
distance-dependent pairwise interaction potentials are capable of
reasonably describing the intermolecular cross correlations of hexane.
This is consistent with the good performance of the MS-CG model in
terms of accurately describing the RDFs. Still, our results
demonstrate that the description of intermolecular interactions as a
function of intramolecular state may assist in alleviating some of the
standard problems experienced with structure-based coarse-graining
(e.g., overly repulsive and over-fitted potentials) while providing a
straightforward approach for characterizing the environment dependence
of CG interactions. Finally, we observe a computational cost of
running the SH simulation to be a factor of 2.0 larger than the
standard CG simulation with the iter-gYBG potentials (i.e., 6.6
ms/step and 3.3 ms/step, respectively, for a box of 267 molecules).

\subsection{Surface transferability}

While our previous study,~\cite{Bereau:2018} as well as the results so
far, highlighted the improved accuracy of conformational surface
hopping over traditional CG structure-based schemes for a single
system or thermodynamic state point, this section explores prospects
of transferability. Without transferability, a new potential would be
required for each new state point in order to reproduce some target
observable. At the other end of the spectrum, excellent
transferability implies that the change in a thermodynamic parameter,
for instance temperature, results in the appropriate change in the
target observable without adjusting the potential. Here we work in an
intermediate, \emph{weaker} transferability regime: We carry over
identical conformational surfaces, and reparametrize their state
probabilities (i.e., the prefactor or weight for each surface). This
approach emphasizes how conformational basins may be shared between
state points, while allowing an adjustment in the overall probability
of that state in a restricted way (i.e., without changing the
corresponding potential). We focus on two aspects: temperature and
chemical composition.

\subsubsection{Temperature and compositional variation}
\label{sec:tempchemtransf}

\begin{figure*}[htbp]
	\begin{center}
		\includegraphics[width=0.9\linewidth]
		{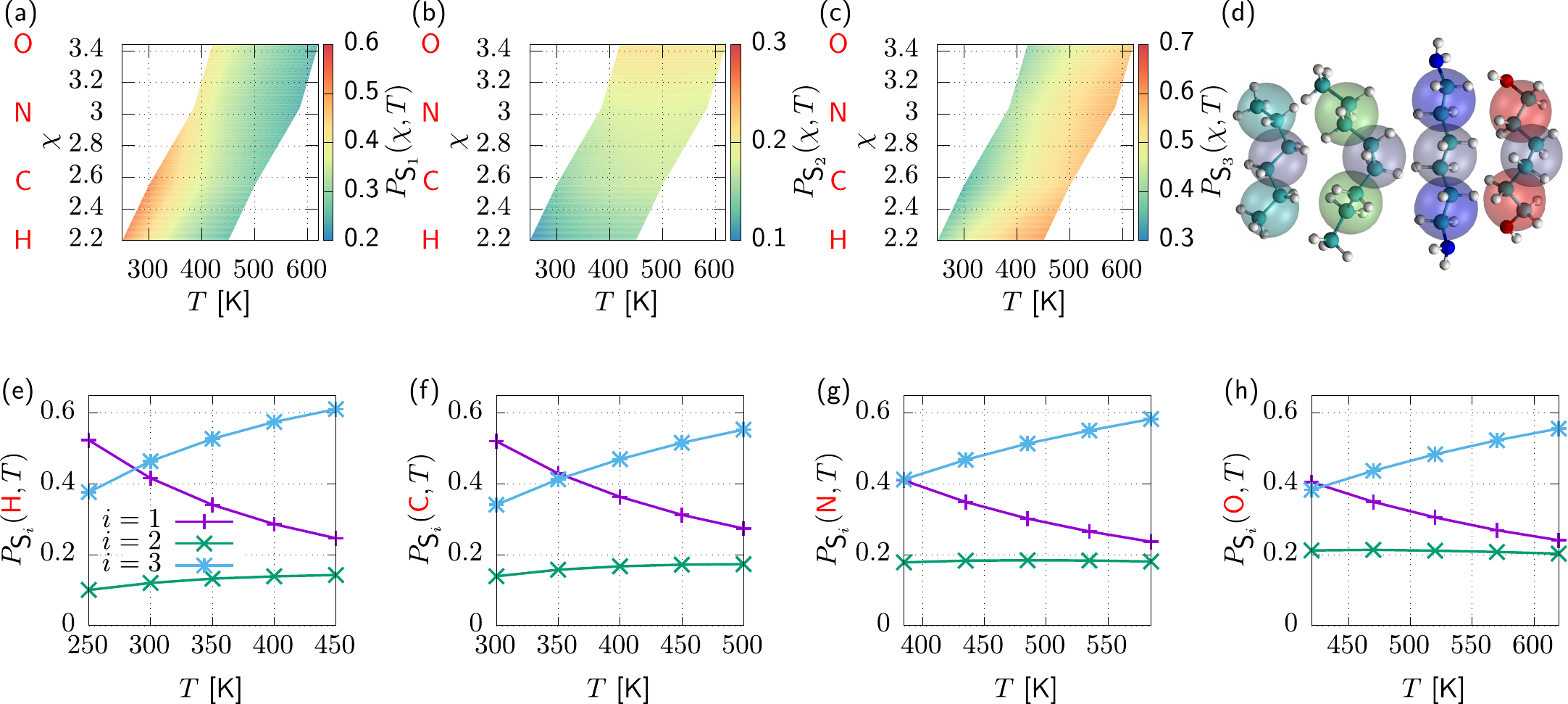}
		\caption{Temperature and compositional transferability. State
			probabilities of the three surfaces, (a) $P_{\textup{S}_1}$,
			(b) $P_{\textup{S}_2}$, and (c) $P_{\textup{S}_3}$ as a
			function of the electronegativity of the substitution atom,
			$\chi$, and temperature $T$. Note the different ranges for the
			color-coding. Substitution atoms H, C, N, and O corresponding
			to hexane, octane, hexanediamine, and hexanediol,
			respectively, are highlighted in red; molecules depicted in
			(d). The different CG bead colors denote the different
			underlying chemical compositions. (e--h) One-dimensional
			projections highlight the smooth temperature dependence of the
			state probabilities. The CG SH individual force fields were
			all constructed from octane, while only tuning the state
		probabilities for each compound separately.}
		\label{fig:twodimtransf}
	\end{center}
\end{figure*}

We consider a set of 3 molecules that are chemically similar to
hexane: octane, hexanediamine, and hexanediol. They correspond to the
same alkane backbone with different terminal substitutions of a methyl
hydrogen on each end: carbon, nitrogen, and oxygen, respectively (with
appropriate saturation), as shown in Fig.~\ref{fig:twodimtransf}d.
Fig.~\ref{fig:twodimtransf} shows the variation of the state
probabilities as a function of both chemical composition and
temperature. The former is described via the electronegativity
parameter $\chi$ of the substitution atom H, C, N, and O corresponding
to hexane, octane, hexanediamine, and hexanediol, respectively. While
we do not provide a formal justification for the use of $\chi$, it is
motivated by the change in the electron density in the terminal
substitutions considered. $\chi$ offers a convenient proxy to describe
the change in chemical composition through a continuous variable.
Further, we have observed monotonic changes of our results with
respect to $\chi$. We will see below that $\chi$ offers a convenient
parameter for scaling the non-bonded interactions.

Panels (a--c) of Fig.~\ref{fig:twodimtransf} show a two-dimensional
projection of the state probabilities for each conformational surface:
$P_{\textup{S}_1}(\chi, T)$, $P_{\textup{S}_2}(\chi, T)$, and
$P_{\textup{S}_3}(\chi, T)$. Panels (e--h) show one-dimensional
projections, highlighting the smooth---almost linear---temperature
dependence. The most significant difference between the three surfaces
is their range of state probabilities: larger for S$_1$ and S$_3$,
while smaller for S$_2$. Surface S$_1$ varies significantly with
respect to both parameters, S$_2$ is more sensitive to composition,
and S$_3$ varies mostly against temperature. Their unique behavior
sheds light on the conformational basin they represent: for instance,
the population of S$_2$ is sensitive to the chemistry, but its low
temperature dependence suggests an enthalpic stabilization. On the
other hand, S$_3$ is rather insensitive to the chemistry, but
significantly varies with temperature. While this could mean that
S$_3$ is stabilized by entropy, we also note that as the fallback
surface it amounts to a collection of different conformational basins.
$P_{\textup{S}_3}(\chi, T)$ varies remarkably little with respect to
chemical composition, given its heterogeneous nature. In what follows
we explore to what extent these smooth variations of the state
probabilities can be leveraged to extend the range of applicability of
a set of force-field surfaces to different state points.

\subsubsection{Temperature transferability}

\begin{figure}[htbp]
	\begin{center}
		\includegraphics[width=0.90\linewidth]
		{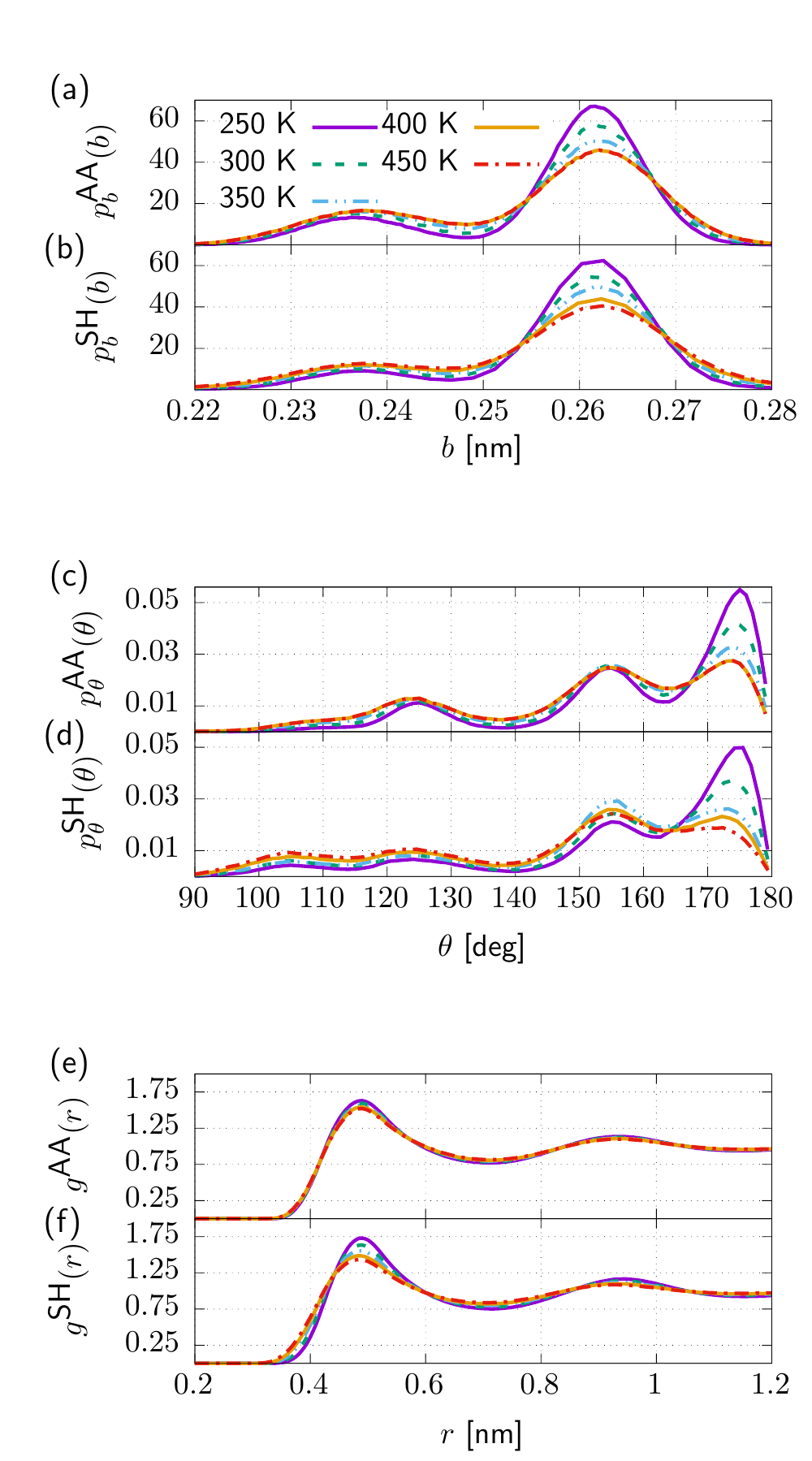} \caption{Temperature
			transferability. Comparison of (a-b) bond and (c-d) bending-angle
			properties of gas-phase hexane between AA and CG resolutions.
			(e-g) Comparison of the RDFs of the CT--CT interactions in the
			liquid phase. The CG SH individual force fields were all
			constructed from the state point at $T=300$\,K, while only tuning
		the state probabilities for each temperature separately.}
		\label{fig:temptransf}
	\end{center}
\end{figure}

We first explore surface transferability across temperature. 
Starting from the three conformational surfaces obtained from
reference AA simulations at $T=300$\,K (see Sec.~\ref{sec:hex_vac}),
we retain these surfaces and only tune the state probabilities to
transfer to the other temperatures $T=\{250,350,400,450\}$\,K. A
comparison of the bond and bending-angle distributions generated by
the AA and SH models are shown in Fig.~\ref{fig:temptransf} (a-d). The
distributions show similar features as found in Fig.~\ref{fig:hexvac},
monotonically evolving as a function of temperature. In particular, we
find a strong temperature dependence of the long bond ($b \approx
0.26$~nm) and the longest angle ($\theta \approx 170^\circ$), while
the other features show virtually no temperature dependence.
Fig.~\ref{fig:temptransf} (e-f) also presents the CT--CT RDFs, which
show reasonable agreement, although the SH distributions are somewhat
too temperature dependent. In comparison to standard transferability
properties, we note that the iter-gYBG model parametrized at
$T=300\,$K and extrapolated to the other state points leads to similar
performance for the one-dimensional distributions (Figs.~S8 and S9).
This is consistent with previous studies that have demonstrated that
temperature-dependent, often linearly-scaled, interactions are
necessary for accounting for the entropic contributions to the
effective potentials.~\citep{Qian:2008,Farah:2011,Lebold:2019b,
Lebold:2019c, Jin:2019} However, the SH model really shines when
considering the description of cross correlations involving the
bending angle (Figs.~\ref{fig:hexvac}, S11 and S12), which standard
parametrizations cannot reproduce.

Our weak transferability scheme offers an accurate reproduction of the
distribution functions for all temperatures, despite the use of a
\emph{single} set of conformational surfaces. The results strongly
suggest a large overlap in conformational space between the
temperatures, adequately captured by retaining the conformational
surfaces and simply adapting the state probabilities to each state
point. We defer a deeper analysis of the temperature dependence of the
state probabilities to Sec.~\ref{sec:tempchemtransf}.

\subsubsection{Compositional transferability}

Beyond the transfer of force fields across temperatures, we now turn
to the more challenging case of compositional transferability---across
chemistry.  We first assessed the transferability of surfaces in the
gas phase, by employing the surface definitions obtained from octane
to each of the other molecules. In this case, all molecules were
simulated at $T=300$\,K. We note that hexane stands as an outlier in
the set of compounds, due to its absence of heavy atoms beyond the six
carbons. The impact of this difference will be illustrated below.

\begin{figure}[htbp]
	\begin{center}
		\includegraphics[width=0.90\linewidth]
		{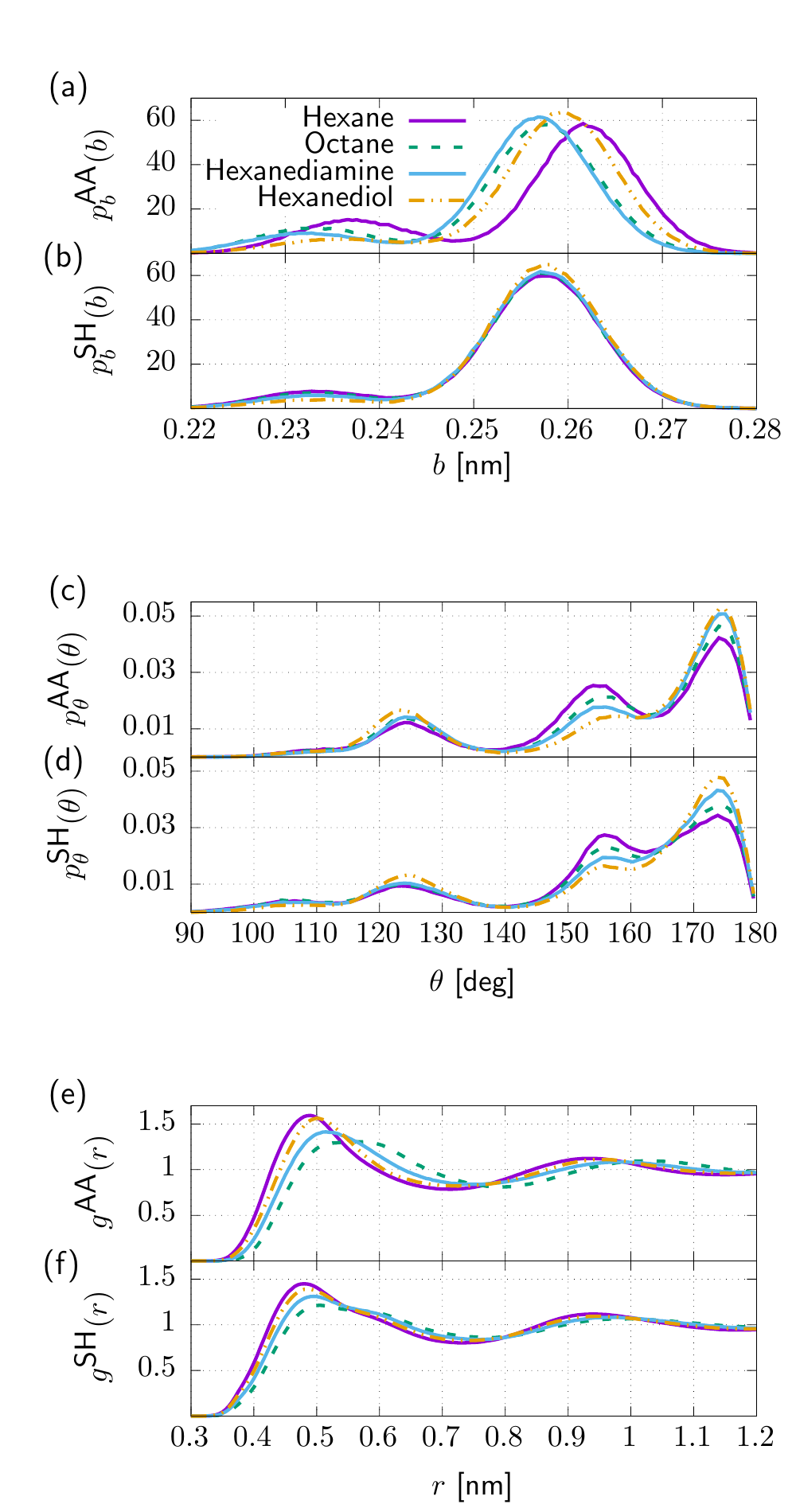}
		\caption{Compositional transferability. Comparison of (a-b) bond
			and (c-d) bending-angle properties of gas-phase hexane, octane,
			hexanediamine, and hexanediol between AA and CG resolutions. (e-f)
			RDFs of the CT--CT interactions in the liquid phase at each
		reference temperature.}
		\label{fig:chemtransf}
	\end{center}
\end{figure}

Panels a and b of Fig.~\ref{fig:chemtransf} show a comparison of the
bond distributions generated by the AA and SH models for the four
molecules. All curves display overall similar behavior. Most
strikingly, we observe a shift in the reference AA distributions:
hexane shows its largest peak at larger values of $b$, while the
others are shifted to lower values, by up to 0.5~\AA. The reason for
this shift is due to our choice of mapping: for consistency reasons we
have kept the terminal CG bead defined as the center of mass of the
two same carbons on the chain. Because of sterics, the presence of
heavy atoms in octane, hexanediamine, and hexanediol have pushed these
carbons slightly inward, resulting in the shifts observed in
Fig.~\ref{fig:chemtransf}a. The interesting aspect here is the impact
this has on our CG models: the use of a single set of surfaces will
necessarily collapse all CG curves, only allowing for vertical shifts
(by varying state probabilities). Interestingly we see little to no
such vertical shifts, unlike what we had observed for temperature
variations (Fig.~\ref{fig:temptransf}).

The bending-angle distributions show variations more in line with the
above-mentioned temperature-variation study: all curves show similar
behavior, i.e., peaks at the same places, with only variations in
their heights. The results are different than the temperature
variation with respect to the relative height differences between
peaks: while varying $T$ led to strong variations in the largest peak,
it had virtually no effect on the second. In contrast, here we observe
variations of similar magnitude between these two peaks. This
strengthens the idea that a local change in chemical composition can
be associated to a perturbation of the conformational space, akin to
changing temperature. However, the local changes between peaks
indicate that alterations occur at a more local level than an overall
temperature rescaling. As a result, it would seem unlikely to reach
compositional transferability of CG force fields by merely scaling it
by a global prefactor. 
Thus, the SH models offer a useful compromise between a limited
prefactor rescaling and state-point dependent potentials, and
highlight the overlap in conformational space of the different
molecules.

We also assessed chemical transferability in the liquid state. We
first directly transferred the non-bonded force field for octane,
while adjusting the state probabilities as described above. Each SH
force field was probed at a distinct temperature $T_{\rm ref}$.
$T_{\rm ref}$---corresponding to 300, 350, 435, and 470\,K for hexane,
octane, hexanediamine and hexanediol, respectively---was chosen to lie
in approximately the same location with respect to the liquid phase
existence for each molecule. This simple transfer of non-bonded
interactions resulted in an underestimate of the changes in the CT--CT
and CT--CM RDFs and an overestimate of the changes in the CM--CM RDF,
while providing a good description of the intramolecular distributions
(see Figs.~S15 and S16). The discrepancies in the RDFs are not
surprising, as we expect that the non-bonded interaction strengths
associated with the CT bead should change as a function of chemistry.
To test the impact of such changes, we performed a simple scaling of
the octane non-bonded interactions. In particular, we applied a
scaling factor to each of the CT--CT potentials equal to the ratio of
electronegativity values of the corresponding substituted atoms:
$U_\textup{{\it M};CT--CT} = (\chi_\textup{{\it M}} /
\chi_\textup{octane}) \, U_\textup{octane;CT--CT}$, where ${\it M} = $
hexane, hexanediamine, or hexanediol. Similarly, the CT--CM potentials
were scaled by the square root of the same ratio (assuming an
effective geometric mean combination rule): $U_\textup{{\it M};CT--CM}
= (\sqrt{\chi_\textup{{\it M}} / \chi_\textup{octane}}) \,
U_\textup{octane;CT--CM}$. The original octane CM--CM interactions
were employed without adjustment. The full set of scaled potentials
are presented in Figs.~S13 and S14.

Remarkably, as shown in panels e and f of Fig.~\ref{fig:chemtransf},
this heuristic scaling of potentials along with the adjusted state
probabilities results in an accurate description of the local CT bead
packing as a function of changes in chemistry, despite employing a
single set of surfaces for the molecules. The accuracy of the CT--CM
RDFs are also improved (relative to the non-scaled SH model) while
retaining a good description of the intramolecular distributions,
although the discrepancy in the CM--CM RDF is somewhat exacerbated
(Fig.~S16). We note that the absolute accuracy of all CG CM--CM RDFs
(i.e., also for the MS-CG and iter-gYBG models) is slightly degraded
due to the challenging representation applied to the non-hexane
molecules, as demonstrated in detail in the SI. Of all the molecules,
hexane yields the largest discrepancies. Its smaller excluded volume
relative to the other molecules represents a larger change in
conformational space: a mere transfer of the conformational surfaces
along with a variation of the non-bonded potentials does not suffice.
These results illustrate the link between shared conformational
surfaces and distance in chemical space.



\section{Conclusions}

This work extends our previous presentation of the coarse-grained (CG)
conformational surface-hopping (SH) methodology: analogous to
switching between different electronic states, we define one force
field per conformational basin and hop between
them.~\cite{Bereau:2018} Each force field is parametrized by
applying force matching (i.e., the MS-CG method) while using only
configurations from the corresponding basin. Our illustration of the
method using a toy example highlights the benefits of enforcing a set
of target state probabilities, which avoids possible errors due to an
inaccurate description of the global surface at the barrier between
two conformational states. While the SH models employ standard
molecular mechanics interaction functions in the Hamiltonian, the
focus on reproducing local properties of each surface results in
increased accuracy relative to standard models. The results are
particularly striking for the gas-phase properties of a three-bead
hexane representation: the correlations between bond and bending
angle, notoriously problematic for the MS-CG method, are accurately
represented with the SH approach while employing only three surfaces.

We have also presented an extension of the SH method to intermolecular
interactions: conformational surfaces are defined based on the
intramolecular state of the molecule, while the intermolecular
interactions depend on the pair of surfaces involved. For instance,
the 3-surface model for hexane consists of two distinct bead types,
which corresponds to a total of 18 unique interactions (i.e., 6
interactions for each pair of bead types). The resulting SH models
retain an accurate description of the local packing, while also
demonstrating slight improvements in the RDFs compared with the MS-CG
model. Perhaps more interestingly, an assessment of the SH potentials
demonstrated promising properties with respect to the other
structure-based potentials. In particular, the SH potentials tended to
be more attractive with a single local minimum, counteracting two
common problems with structure-based models: ($i$) the underestimation
of the cohesive energy in liquids and ($ii$) an over-fitting of the
features at the state point of parametrization.

We further investigated the capabilities of the SH models by examining
their transferability properties. We focused on a so-called
``weak-transferability regime,'' in which one state point determines
the surface definitions; these surfaces are then transferred to other
state points, while adjusting their state probabilities (i.e., the
prefactor or weight for each surface). In particular, we considered the
transfer of state definitions across both temperature and chemical
composition. In the latter case, where the strength of the
interactions are expected to change as a function of chemistry, the
use of the electronegativity parameter, $\chi$, provided a useful
proxy to scale the non-bonded interactions. Our results demonstrate
that the SH models not only accurately describe the trends in the
intramolecular distributions, which are largely reproduced with
traditional models, but also better represent intramolecular cross
correlations throughout the liquid state. The SH approach demonstrates
similar results with respect to the description of local packing as a
function of temperature for the molecules considered, but slightly
overestimates the temperature dependence of the RDFs. It would be
interesting in this context to explore the entropic contributions to
the SH potentials.~\cite{Lebold:2019b, Lebold:2019c, Jin:2019} The
investigation of chemical transferability focuses on terminal
substitutions via the comparison between hexane, octane,
hexanediamine, and hexanediol.  Notably, we find limitations in
modeling the bond distributions, as the substitution of hydrogens to
heavy atoms (i.e., moving from hexane to one of the other three
molecules) shifts the distribution. Aside from this limitation, the
tuning of individual state probabilities appears to be a promising
framework for considering the construction of CG models that are not
restricted to one state point, but rather applicable to a neighborhood
of thermodynamic parameters and chemical compositions. An almost
linear variation of the state probabilities is observed across both
temperature and electronegativity, making it straightforward to
interpolate across this set of state points. Here we did not intend to
make predictions across chemical space, but rather explore to what
extent transferability via (only) changes in thermodynamic variables 
can be facilitated through their impact on individual surfaces.
The approach highlights
overlaps of conformational basins across neighborhoods of chemical
space. We foresee the weak-transferability regime brought forward here
to be of use when parametrizing not just one reference simulation, but
collections of state points or compounds. This will be of use in the
context of parametrizing CG models across subsets of chemical space.

Finally, we stress the conceptual and practical advantage of
parametrizing the SH models using the MS-CG technique. The combined
approach offers an enhanced capability to describe complex cross
correlations between degrees of freedom that arise additively in the
Hamiltonian, while using a \emph{direct} inverse parametrization
scheme. Since the MS-CG method results in errors whenever the AA cross
correlations represent an inappropriate proxy for the cross
correlations of the resulting CG model, the approach provides an
automatic validation of the surface definitions. In other words, if
there remain cross correlations within a single surface that cannot be
reproduced by a molecular mechanics force field, errors will likely
appear in the description of the modes along each distribution
function corresponding to the inadequate surface. Moreover, the
potentially large number of force fields---up to one per
conformational basin---can be derived independently, an aspect that
would not be straightforward using iterative methods.

\section*{Acknowledgments}
The authors thank Oleksandra Kukharenko and Christoph Scherer for
critical reading of the manuscript. This work was partially supported
by the Emmy Noether program of the Deutsche Forschungsgemeinschaft
(DFG).

\bibliography{biblio,references_MPIP,references_PSU} 

\end{document}